\documentclass[twocolumn,showpacs,preprintnumbers,amsmath,amssymb]{revtex4}

\usepackage{graphicx}
\usepackage{dcolumn}
\usepackage{bm}
\usepackage{amsmath}

\begin{document}


\title{Proposal for testing Einstein's moon using three-time correlations}

\author{Toshiyuki Fujii$^1$}
\author{Munehiro Nishida$^{2}$}
\author{Noriyuki Hatakenaka$^{1}$}%

\affiliation{$^1$Graduate School of Integrated Arts and Sciences, Hiroshima University, 
Higashi-Hiroshima, 739-8521, Japan
}%
\affiliation{$^2$Graduate School of Advanced Sciences of Matter, Hiroshima University, Higashi-Hiroshima,
739-8530, Japan.
}%

\date{\today} 

\begin{abstract}
Quantum mechanics has predicted  many counterintuitive phenomena in daily life, 
and has changed our view of the world. Among such predictions, 
the existence of a macroscopic object in superposition is especially unbelievable. 
As Einstein asked, ``Do you really believe that the moon exists only when you look at it?". 
However, recent experimental results on a mesoscopic scale will ultimately require us 
to dismiss  commonsense so-called macroscopic reality. 
Leggett and Garg applied the Bell scheme for testing local realism 
to the time evolution of a macroscopic two-state system, 
and proposed a temporal version of the Bell inequality (the Leggett-Garg (LG) inequality) 
for testing macroscopic realism. However, as with the Bell inequality, 
the statistical approach behind this scheme may be less effective in showing clear incompatibility. 
Here we propose a temporal version of  the Greenberger-Horne-Zeilinger (GHZ) scheme 
without statistical treatment for testing Einstein's moon using three-time correlations.  
\end{abstract}

\maketitle
\begin{figure}
\begin{center}
\includegraphics[width=8.5cm,clip]{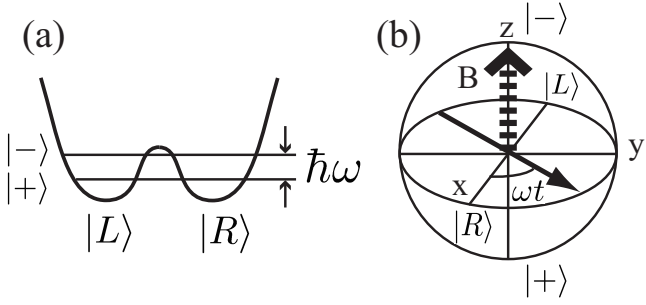}
\caption{\label{tghzsys} (a) The double-well system and (b) its spin representation (Bloch sphere)D
The $z$ axis of the Bloch sphere in  (b) represents the bonding state 
$|+\rangle =  ( |R\rangle  +|L\rangle )  / \sqrt{2}$
and the antibonding state 
$|-\rangle = ( |R\rangle  -|L\rangle )  / \sqrt{2}$, 
and the $x$ axis represents $|R\rangle $ and $|L\rangle $. 
The Pauli operators are defined as 
$
\hat \sigma_x =  |R\rangle \langle R | -|L\rangle \langle L |, 
\hat \sigma_y = i (   |L\rangle \langle R | -|R\rangle \langle L | ),$ and
$
\hat \sigma_z =- (|R\rangle \langle L |  +|L\rangle \langle R | ). 
$ 
The Hamiltonian for the double well system in the Pauli representation is expressed as 
$
\hat H= \hbar \omega \hat \sigma_z /2  
$
where $\hbar \omega $ is the energy separation between the states $|\pm \rangle$. 
}
\end{center}
\end{figure}

The quantum mechanical formalism has been extraordinarily successful 
in accounting for a wide range of microscopic phenomena, 
while its applicability to macroscopic scales remains  unclear. 
One striking example is Schr\"{o}dinger's dead-and-alive cat \cite{Scat}, 
which serves to demonstrate the apparent conflict between classical and quantum theory 
on a macroscopic scale. It has been examined in superconducting quantum interferometer devices (SQUIDs), 
i.e.,  a superconducting loop interrupted by a thin insulating layer with a small area. 
The supercurrent of about one microampere circulating in the loop, 
which is equivalent to the net magnetic flux threading the loop, 
is a relevant {\it macroscopic} variable characterizing the system. 
Under appropriate conditions, the potential energy of the system is a ``double potential well" 
with minima sufficiently far apart, and so regarded as macroscopically distinct. 
We specify these two states using the dichotomic variable $q =\pm 1 $. 
The ground states in the right and left well correspond to the current circulating 
in the clockwise $(q=+1)$ and  anticlockwise $(q=-1)$ directions, respectively.
According to standard quantum mechanics, 
a coherent oscillation of these counterflowing current states ($|R \rangle $ and $|L\rangle $) occurs 
due to the superposition of the states if the system is suitably decoupled from its environment 
(see Fig. \ref{tghzsys} (a)).  
At the end of the twentieth century, the tail of the cat fabricated by modern technology 
has been caught in superconducting nanocircuits by observing the energy splitting 
caused by the superpositions of macroscopically distinct states \cite{Friedman, Casper}.

However, this is insufficient to  confirm the validity of quantum mechanics on a macroscopic scale. 
Closely related is  Einstein's moon paradox \cite{Einstain}, 
which concerns the notion of objective reality. 
Einstein's argument suggests that the moon (matter) must have a separate reality 
with definite values at all times independent of measurements. 
Leggett and Garg (LG) have challenged this philosophical question 
by moving the discussion to the level of mathematical demonstration and experiments \cite{LG}.  
They first define the notion of reality on a macroscopic scale as {\it macrorealism per se}, 
which is implicit in much of our thinking about the macroscopic world. 
They then derived  Bell-type inequalities in the {\it time} domain 
under two supplemental assumptions {\it induction } and {\it noninvasive measurability}, 
yielding constraints on the behavior of macroscopic systems 
that are incompatible with some predictions of quantum mechanics. 
These inequalities are similar to Bell's inequalities 
for the Einstein-Podolsky-Rosen (EPR) experiment \cite{Bell}. 
In contrast to the Bell test, LG involves successive measurements 
at different times on a {\it single} system. 
The measurement times, $t_i$, play the role of the polarizer settings in the ordinary Bell inequalities. 

Despite of considerable experimental and theoretical efforts 
\cite{Tesche, paz, Huelga, calarco, Ruskov, Jordan, williams}, 
no violation has yet been reported basically due to the problem of ensemble preparation. 
Greenberger, Horne and Zeilinger (GHZ) developed a proof of the Bell theorem without inequalities \cite{GHZ}. 
The entanglement of more than two particles leads to a strong conflict 
between local realism and non statistical predictions of quantum mechanics. 
In this Letter, we propose a new scheme for testing Einstein's moon without statistical treatments 
by combining LG and GHZ ideas, i.e., temporal GHZ scheme. 

Let us consider the measurement of the $q$ value of the system at a certain time $t$. 
One requirement for macrorealistic theories is that any measurements must be noninvasive. 
A negative-result measurement is one possible scheme that has been previously considered \cite{LG, Tesche}. 
Here we employ an orthodox noninvasive measurement \cite{LG}, i.e., a microscopic probe. 
Here it is assumed that the back action of the measurement is too small 
to cause any great change in the consequent motion of the macroscopic system. 
The state of the system is indirectly measured 
via a microscopic object that undoubtedly behaves quantum mechanically 
taking two states labeled as $|0\rangle $ and $|1 \rangle$ (ancilla bit). 
In the first step of the indirect measurement, the ancilla bit probes the system by interaction with it,
the system information is registered by flipping state of the ancilla bit initially 
prepared as $|0\rangle$ if the system has a value $q=+1 $, otherwise keeps it unchanged. 
The ancilla then acts a memory by taking the state $|1 \rangle $ or $| 0\rangle$ 
corresponding to the system value $q=+1$ or $q=-1$, respectively. 
The result of this measurement step is stored tentatively in the ancilla bit. 
Then, the ancilla bit's state is  measured directly in the usual way as the second step.

\begin{figure}
\begin{center}
\includegraphics[width=8cm,clip]{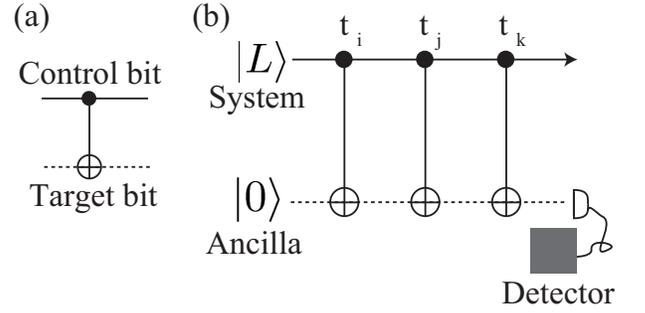}
\caption{\label{tghztest} 
Circuit diagram of (a) a CNOT gate operation and (b) three  successive  CNOT measurements 
together with subsequent readouts of the ancilla bit. }
\end{center}
\end{figure}

The operation in the first step of the indirect measurement is considered 
to be a controlled NOT (CNOT) operation in terms of quantum information technology  
and is represented as a circuit diagram in Fig. \ref{tghztest} (a) \cite{NC}. 
The unitary operator for the CNOT operation can be expressed as 
\begin{equation}
\begin{split}
\hat s &=| R \rangle \langle R| \otimes \hat X +  | L \rangle \langle L| \otimes \hat I \\
& =\frac{1}{2} ( \hat 1 + \hat \sigma_x   ) \otimes \hat X 
+\frac{1}{2} ( \hat 1 - \hat \sigma_x   ) \otimes \hat I
\end{split}
\end{equation}
where $\hat 1 = |R \rangle \langle R | +  |L \rangle \langle L | $ is an identical operator for the system. 
$\hat X$ and $\hat I$ are a flip operator and an identical operator for the ancilla bit. 
$\hat{\sigma}_x$ is a spin operator of the $x$ component defined as 
$\hat \sigma_x =  |R\rangle \langle R | -|L\rangle \langle L |$. 
In the Heisenberg picture, the operator representing the CNOT operation at $t$ is expressed as 
\begin{equation}
\begin{split}
\hat s(t) &=\hat u^{\dagger}( t) \hat s \hat u( t) =\hat R_z(-\omega t) \hat s \hat R_z^{\dagger}(-\omega t) \\
& =\frac{1}{2} \left \{  \hat 1 + \hat \sigma({-\omega t} ) \right\}  \otimes \hat X +\frac{1}{2} 
\left \{ \hat  1 - \hat \sigma({-\omega t})   \right\} \otimes \hat I \label{st},
\end{split}
\end{equation}
where 
$
\hat \sigma (-\omega t) 
= \hat \sigma_x \cos{\omega t } -  \hat \sigma_y \sin{\omega t }
$
and 
the time-translational operator $\hat u(t)$ is equivalent to the rotational operator 
around the $z$ axis:  
\begin{equation}
\hat u(t)= \exp{\left [ -i \frac{\omega t }{2}\hat \sigma_z \right]} = \hat R_z(\omega t).
\end{equation} 
Thus the state evolution of two-level system is regarded 
as the spin precision under a static magnetic field directed towards the $z$ axis.
This shows that time-translational operators in the Heisenberg picture 
play a role of the change of the polarizer angle in a spatial scheme.

Leggett and Garg introduced a two-time correlation function on a single object 
to test the validity of the macrorealistic theory.  
This scheme is essentially a temporal version of the Bell scheme, 
which tests the local realistic theory by using the correlation between two particles, 
and  the same experiment must be repeated many times to obtain the average correlation value 
precisely in statistical treatments.
On the other hand, the GHZ scheme can tests the local realistic theory without statistical treatment 
by using the correlation between three particles. 
Therefore we consider the three-time correlation of a single object 
for testing the macrorealistic theory without statistical treatment.

From  a macrorealistic viewpoint, the system has a definite $q$ value at any time. 
Three CNOT operations then flip the ancilla-bit's state in an odd-number time for $q_kq_jq_i=+1$, 
resulting in $|1\rangle $Cand in an even-number time for $q_kq_jq_i=-1$ with the output $|0\rangle$. 
As a result, the system's state is stored in the ancilla bit after all the operations 
and the output of the ancilla bit represents the product of the system's values  
at each CNOT time, i.e.,  $q_kq_jq_i =\pm 1$. 
This product is the key to our temporal scheme as with the spatial GHZ scheme. 

In a {\it spatial} GHZ test, two of three spin components, say $x$ and $y$, are used 
for each particle of three entangled  particles. 
The correlation measurement is performed for a total of six variables. 
Similarly, a {\it temporal} GHZ scheme requires six different choices of measurement times ($t_i$, $i=1-6$), 
from which three-time combinations such as $(t_1, t_2, t_3)$ are selected for CNOT operations. 
We can assume that the CNOT operations for combinations $(t_1, t_4, t_5), (t_2, t_5, t_6)$, 
and $(t_3, t_4, t_6)$ generate $|1 \rangle $ as the ancilla-bit state without loss of generality. 
In terms of macroscopic realism, this is equivalent to the following three GHZ-like relations:     
\begin{equation}
\begin{cases}
\text{(i)} & q_5q_4q_1 =+1 \\
\text{(ii)}& q_6q_5q_2=+1\\
\text{(iii)}& q_6q_4q_3=+1. \label{tghzrelations}
\end{cases}
\end{equation}
These conditions should simultaneously hold because the  $q_i$ values are predetermined 
at the start of system evolution
whether or not measurements are performed. 
Therefore, the product of these conditions yields   
\begin{equation}
\label{tghzrelations2}
\text{(iv)} \hspace{0.2cm} q_1q_2q_3=+1.
\end{equation}
Here we use the identity for the dichotomic variable  $q_i^2 =1$. 
This is a key relation of a temporal GHZ scheme. 
Any violation of Eq. (\ref{tghzrelations2}) completely rules out macrorealistic theories. 

If the system obeys quantum mechanics, each CNOT operation only results in an entanglement 
between the system and the ancilla bit, 
and then the system does not have a definite $q$ value until the ancilla is detected. 
Therefore, three successive CNOT operations cannot be regarded 
as a measurement of the product $q_kq_jq_i $ from a quantum-mechanical standpoint. 
Thus, there is no clear correspondence between  ancilla bit value and $q_k q_j q_i$. 
However, despite this lack of correspondence, 
the violation of Eq. (\ref{tghzrelations2}) can still be tested if the ancilla state is measured, 
since the condition (iv) implies that  the ancilla state is in $|1 \rangle $ 
from a macrorealistic viewpoint. 
Therefore, the observation of the $|0\rangle $ state for the CNOT combination $(t_1, t_2, t_3)$ 
can be clear evidence for the violation of macrorealism. 
\begin{figure}
\begin{center}
\includegraphics[width=8.5cm,clip]{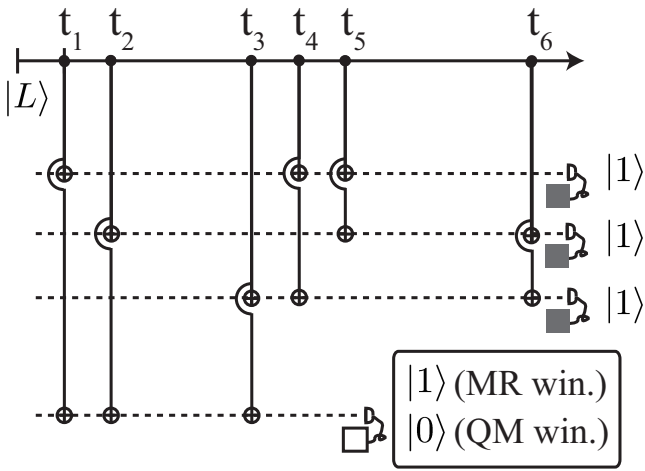}
\caption{\label{tghz}  Circuit representation of a temporal GHZ test consisting of measurements 
at four different combinations of successive three times.}
\end{center}
\end{figure}
We now show that condition \eqref{tghzrelations2} is indeed violated when the system obeys quantum mechanics.
The successive CNOT operations at $t_i$, $t_j$, and $t_k$ are described as 
\begin{equation}
\begin{split}
\hat s(t_k) \hat s(t_j) \hat s(t_i) \hspace{-2cm} & \\
&= \frac{1}{2} \Big [  \hat X \otimes  \left\{ \hat 1 + \hat \sigma (-\omega t_k)\hat \sigma (-\omega t_j) \hat \sigma  (-\omega t_i) \right\} \\
&\hspace{1cm}+  \hat I \otimes  \left\{ \hat 1 - \hat \sigma (-\omega t_k)\hat \sigma (-\omega t_j) \hat \sigma  (-\omega t_i) \right\} \Big]. 
\label{tsokutei}
\end{split}
\end{equation} 
Since the relation  
\begin{equation}
\hat \sigma (-\omega t_k)\hat \sigma (-\omega t_j) \hat \sigma  (-\omega t_i) =\hat \sigma (-\omega (t_k-t_j+t_i)), \label{tseki}
\end{equation}
the product of three operators is formed into a {\it single} spin operator. 
This describes a spin operator rotated clockwise with an angle $\omega (t_k-t_j+t_i)$ 
from the $x$ in the $x-y$ plane of the Bloch sphere. 
The outputs of the ancilla bit become $|1\rangle $  if $t_i $  satisfies the following conditions;
\begin{equation}
\begin{cases}
t_5-t_4+t_1 &=(2k +1) \pi/\omega \\
t_6-t_5+t_2 &=(2l+1 ) \pi/\omega \\
t_6-t_4+t_3 &=(2m+1) \pi/\omega ,  \label{tghzxyy}
\end{cases}
\end{equation}
where $k,l,$ and $m$ are integers. 
In macrorealistic terms, this means situations (i), (ii), and (iii) are  achieved.

Now let us consider the following additional relation for $t_1$, $t_2$ and $t_3$;  
\begin{equation}
t_3-t_2+t_1=\frac{2n\pi}{\omega }, \hspace{1cm } n=0,\pm 1. \cdots, \label{tghzxxx}
\end{equation}
This relation does not change the condition (iv) as long as the relations \eqref{tghzxyy} 
holds in macrorealistic theories, equivalently $|1\rangle $ for the ancilla bit. 
In this case, the operator \eqref{tseki} becomes $\sigma_x $, 
resulting in the ancilla state $|0\rangle $ for the system initial state $|L\rangle $. 
This is completely opposite to the prediction inferred from macrorealistic condition (iv). 
Therefore, we can rule out macrorealistic theories 
from the clear discrepancy as regards the ancilla bit 
in a single run of experiments simultaneously satisfied in \eqref{tghzxyy} and \eqref{tghzxxx}. 
\begin{figure*}
\begin{center}
\includegraphics[width=15cm,clip]{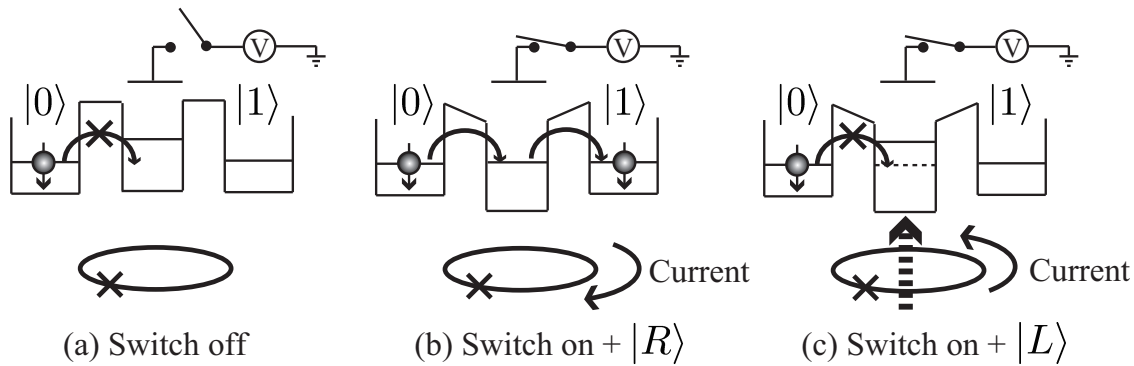}
\caption{\label{nim} An example of the experimental model 
for noninvasive measurements using an electron in a triple quantum dot. 
The $|R\rangle$ and $|L\rangle$ states produce different magnetic fields at the center of the dots, 
leading to different Zeeman shifts, for simplicity, zero for the $|R\rangle$ state. 
(a) A SQUID is energetically decoupled from a triple quantum dot. (No measurement) 
The external bias voltage aligns the levels in the entire system. 
(b) A resonant tunneling occurs due to no extra energy when the SQUID is in the $|R\rangle$ state. 
This flips the electron's state. 
(c) A tunneling is prohibited due to an extra Zeeman energy caused by the $|L\rangle$. 
The electron remains in the same dot. 
The case (b) together with (c) show the CNOT operation. 
}
\end{center}
\end{figure*}
There are a number of solutions that  satisfy the above four conditions of the six successive times. 
One example is 
$(t_1,t_2,t_3,t_4,t_5,t_6)= \frac{\pi }{\omega }(\frac{1}{2},1,\frac{5}{2},3,\frac{7}{2},\frac{11}{2})$
for $(k,l,m,n)=(0,1,2,1)$ as shown in Fig. \ref{tghz}.

A possible experimental setup that can be performed with current technology is shown in Fig. \ref{nim}. 
A SQUID (system) is inductively coupled to the center quantum dot with a giant g-factor \cite{Zhang} 
in a quantum nanostructure with three quantum dots. 
An electron probes the SQUID state through a tunneling between left and right dots 
since the energy level configuration in the center dot depends on the SQUID state. 
Fig. \ref{nim} shows an example of the CNOT operation in this system. Three CNOT operations designed in three different definitely planned times provide a final result in question. 

In summary, we have proposed a nonstatistical scheme 
for testing the validity of macrorealistic theories using three-time correlations 
on a single object based on the GHZ scheme. 
This proposal allows us to clarify directly the incompatibility 
between quantum mechanics and a realistic framework, 
showing clear discrepancies from realistic predictions 
from a series of definite measurement outputs without statistical treatment.

We thank S. Kawabata for valuable discussions. 
This work was supported in part by KAKENHI (Nos. 18540352,  20540357, and 195836) from
MEXT of Japan.

\end{document}